\documentclass[11pt]{article} 
\usepackage{moriond,epsfig} 
\usepackage{tabularx} 
\usepackage{longtable} 
 
\def\Journal#1#2#3#4{{#1} {\bf #2}, #3 (#4)} 
 
\def\AA{{\em Astronomy \& Astrophysics}} 
\def\AAS{{\em Astronomy \& Astrophysics Suppl. Ser.}} 
\def\AJ{{\em The Astronomical Journal}} 
\def\APJ{{\em Astrophysical Journal}} 
\def\MNRAS{{\em Mon. Not. R. Astron. Soc.}} 
 
\def\be{\begin{equation}} 
\def\ee{\end{equation}} 
\def\bea{\begin{eqnarray}} 
\def\eea{\end{eqnarray}} 
 
\begin{document}
\begin{flushright}{\bf LAL 99-70}\\
November 1999
\end{flushright}

\vspace*{4cm}
\title{\large A NEARBY SUPERNOVAE SEARCH: EROS2}

\author{{\bf N. REGNAULT} \footnote{on behalf of the EROS2 Collaboration.}}
\address{{\bf \large Laboratoire de l'Acc\'el\'erateur Lin\'eaire,}\\
IN2P3-CNRS et Universit\'e de Paris-Sud, BP 34,  F-91898 Orsay Cedex} 
 
\maketitle\abstracts{ 
Type Ia supernovae (SNIa) have been  used as 
approximate standard candles to measure cosmological parameters  
such as the Hubble constant and the deceleration parameter. 
These measurements rely on empirical correlations between 
peak luminosities and other features that can be observed  
in the supernovae spectra and their light curves.  
Such correlations deserve further study since they have been  
established  from small samples of nearby SNIa. 
Two years ago, the EROS2 collaboration launched an automated search 
for supernovae with the 1m Marly telescope operating at La Silla.  
In all, 57 SNe have been  discovered in this EROS2 search 
and spectra have been  obtained for 26 of them. We found that  
75\% were of type Ia and 25\% of type II. Using this sample, a preliminary SN   
explosion rate has been obtained. Our most recent observation campaign  
took place in February and March 99. It was performed  in the framework  
of a large consortium  led by the {\em Supernova Cosmology Project}.  
The aim of this intensive campaign was 
to provide an independent set of high quality light curves  
and spectra to study systematic effects in the measurement of  
cosmological parameters. We will briefly describe our search 
procedure and present the status of our ongoing analysis. } 
 
\section{Type Ia supernovae and cosmology}  
 
 
Supernovae are classified in different subtypes, according to their  
spectral features. Type Ia supernovae (SNIa) are believed to be  
explosions of carbon-oxygen white dwarfs. SNIa progenitors are  
likely to be binary systems, composed of a red giant and an old  
C/O white dwarf. The latter accretes matter from its companion  
until it reaches the Chandrasekar mass ($\sim 1.4 M_\odot$),  
and then becomes unstable.  
This process leads to the total thermonuclear explosive burning 
of the white dwarf. Thus, the total energy released should be nearly constant  
from one SNIa to another. These objects may therefore be used as  
standard candles.  
Indeed, photometric and spectroscopic studies have shown that SNIa  
compose an homogeneous sample and their peak magnitudes present a small scatter  
($\sim 20\%$) in all colors. Furthermore, these  
objects are very luminous --- they have been detected  
up to redshifts $z \sim 1.2$ (Aldering {\em et al.}~\cite{ald}).  
Thus they constitute powerful cosmological distance indicators.  
 
 
It has been shown  that the absolute maximum luminosities of SNIa correlate with other  
observables, like the post maximum decline rate, the color at maximum,  
the SN spectral features, or the galaxy type. When corrections based on such correlations  
are applied, the relative dispersion of the peak luminosities of SNIa can be reduced to 10\% (Hamuy {\it et al.}~\cite{ham}).  
 
Measurements of the cosmological parameters $H_0$, $\Omega_0$ and  
$\Lambda$ have been made by analysing the apparent peak magnitude  
versus redshift relation (Perlmutter {\it et al.}~\cite{perl},  
Schmidt {\it et al.}~\cite{sch}).  
These analyses rely heavily on the standardization procedure outlined 
above. For example, the evidence for a non zero $\Lambda$ arises from  
a 20\% flux decrease with respect to a ($\Omega=0.2$) universe,  
which is comparable to the intrinsic luminosity spread.  
%
However, our SNIa knowledge is based on few objects, namely the 17 SNIa  
($z<0.1$) discovered before maximum during the Calan-Tololo search. This  
is why a number of nearby SN searches have been launched in order  
to increase the set of well sampled SNIa and study further the  
standardization corrections mentionned above.  
 
Supernovae rates as a function of redshift are a useful tool for  
studying the star formation history, or constraining the galactic chemical  
evolution scenarios. While probing the stellar evolution, they also bring valuable  
information on the SNIa progenitor system, and allow us to  
get a better insight into the physics processes involved in these events.  
SNIa rates have been measured at low redshift (see {\em e.g.} Cappellaro 
{\it et al.}~\cite{cap}~) with SNe discovered using photographic plates,  
and at high redshift ($z \sim 0.4$), with automatic subtraction of  
CCD images by Pain {\it et al.}~\cite{pain}. EROS2 has obtained the first  
determination of the SNIa rate at $z \sim 0.15$ (Hardin {\it et al.}~\cite{delph}).  
 
\section{The EROS2 nearby supernovae search}  
 
The EROS2 experiment is mainly devoted to the search for  
microlensing events towards the Magellanic clouds, and towards the Galactic  
bulge and disk. For this purpose, the collaboration operates  
a 1 meter telescope, installed at the {\em European Southern Observatory} 
of La Silla (Chile). This instrument was specially refurbished and automated  
in view of a microlensing survey. It is equipped with a dichroic beam splitter  
and two cameras to take images simultaneously in two wide  
pass-bands. Each camera comprises a mosaic of 8 $2k \times 2k$ thick CCD's,  
covering a field of view of $0.7^o(\alpha) \times 1.4^o(\delta)$  
with a pixel size of 0.6 arcseconds.  
 
Since this setup is particularly well suited for discovering supernovae  
at $z \sim 0.05-0.2$, the EROS2 collaboration launched in 1997 a systematic  
nearby SN search aimed at the measurement of the  
nearby SN explosion rates and a detailed study of the correlations between  
the SNIa light curve shapes and their peak absolute luminosities.

\subsection{The search strategy} 
 
Our SN search technique consists in comparing an image of a given field 
with a {\em reference image} of the same field taken  
two or three weeks before. For this purpose, we subtract the {\em reference 
frame} from the {\em search frame}, after a geometric and a photometric alignment,  
and a matching of the seeing. We then perform an object detection  
on the subtracted frame.  
Genuine candidates are selected among these objects by applying cuts  
tuned with a Monte-Carlo simulation, in order to reject variable stars,  
asteroids and subtraction artifacts. Finally, a visual scan allows  
us to eliminate the last spurious candidates.  
 
\subsection{The first stage : 1997-1998} 
 
During the first two years, 7 search campaigns have been conducted.  
We monitored fields from both celestial hemispheres. In order to avoid  
dust absorption they were chosen far from the Galactic plane. 
During these first searches, 35 SNe have been discovered.  
Spectra could be obtained for 10 of them with the ESO 3.6m and  
the ARC 3.5m telescopes. 7 of the SN were of type Ia, 1 of type Ic and  
2 of type II. Using this first sample, a preliminary SNIa rate  
at $z \sim 0.15$ has been obtained (see section \ref{sec:snrate}).

\subsection{A worldwide SNIa search campaign} 
 
In the spring of 1999, we participated in a worldwide search campaign 
\footnote{Involving the following 9 groups :  
{\bf The Nearby Galaxies SN Search Team} ($\dagger$) (Strolger, Smith {\em et al.}),  
{\bf EROS2} ($\ddagger$) (Spiro {\em et al.}),  
{\bf KAIT} (${\dagger\dagger}$) (Filippenko {\em et al.}),  
{\bf The Mount Stromlo Abell Cluster Supernovae Search} (${\ddagger\ddagger}$) (Schmidt, Germany \& Reiss),  
{\bf NEAT} ($\bot$) (Helin, Pravdo \& Rabinovitz),  
{\bf QUEST} ($\top$) (Schaefer {\em et al.}),  
{\bf SpaceWatch} ($\ast$) (McMillan \& Larsen),  
{\bf The Tenagra Observatories} ($\diamond$) (Schwartz),  
and {\bf The Wise Observatories Supernovae Search} ($\bigtriangledown$) (Gal-Yam {\em et al.}).},  
led by the {\em Supernovae Cosmo\-logy Project} and coordinated  
by Greg Aldering (SCP). The search involved 9 groups listed below.  
 

EROS2 discovered a subset of 16 SN among the 41 supernovae found in 
this campaign. Among them, 19 (7 from EROS2) turned out to be of type 
Ia, discovered near maximum. An overview of the follow-up data for 
each SN can be found in table \ref{tab:follow-up}.  Photometric and 
spectroscopic data from both these SN together with discoveries 
announced in the same period in IAU circulars are currently being 
analysed.

\begin{table}[h] 
\caption{An overview of the photometric follow-up of the SNIa discovered during  
the Spring 1999 SN search.\label{tab:follow-up}} 
\vspace{0.4cm} 
\begin{center} 
{\scriptsize 
\begin{tabular}{|c|cccccccccc|} 
\hline 
SN & 99aa 
   & 99ao(${\ddagger\ddagger}$) 
   & 99ac(${\dagger\dagger}$) 
   & 99af($\ddagger$) 
   & 99ar($\dagger$) 
   & 99at($\bot$) 
   & 99au($\dagger$) 
   & 99av($\dagger$) 
   & 99aw($\dagger$)  
   & 99ax($\bigtriangledown$)\\ 
   &      &      &      &      &      &      &      &      &      &    \\  
U  & 22   &  17  & 23   &  8   &  6   &  7   &  10  &  15  &  19  & 13 \\  
B  & 27   &  16  & 27   &  9   & 11   &  8   &  13  &  19  &  18  & 10 \\  
V  & 27   &  16  & 28   & 11   & 11   &  9   &  12  &  18  &  19  & 16 \\  
R  & 27   &  15  & 27   & 10   & 11   &  9   &  12  &  17  &  19  & 15 \\  
I  & 22   &  14  & 19   &  7   &  5   &  6   &  12  &  11  &  13  &  9 \\  
\hline  
   & 99be($\ast$) 
   & 99bf($\ast$) 
   & 99bh($\dagger\dagger$) 
   & 99bi($\ddagger$) 
   & 99bk($\ddagger$) 
   & 99bm($\ddagger$) 
   & 99bn($\ddagger$) 
   & 99bp($\ddagger$) 
   & 99bq($\ddagger$) 
   & 99by($\dagger\dagger$) \\ 
   &      &      &      &      &      &      &      &      &      &      \\  
U  &  7   &  --  &  7   &  9   &  5   &  1   &  5   &  10  &  1   &  9   \\  
B  & 12   &  4   & 12   &  9   &  7   &  7   & 11   &  10  &  4   & 15   \\  
V  & 12   &  5   & 11   &  8   &  12  &  6   &  9   &  10  &  5   &  9   \\  
R  & 11   &  3   & 12   &  9   &  10  &  4   &  7   &  11  &  2   &  9   \\  
I  &  8   &  1   &  8   &  7   &  8   &  1   &  6   &   6  &  2   &  9   \\  
 
\hline 
\end{tabular} 
} 
\end{center} 
\end{table}

\section{A first measurement of the SNIa rate at $z \sim 0.15$}\label{sec:snrate} 
 
Our preliminary determination of the SNIa rate at $z \sim 0.15$ relies  
on a sample of type Ia supernovae discovered during 2 search campaigns 
led in October and November 1997. 120 square degrees have been covered,  
8 supernovae discovered. Among them, 4 were of type Ia.  
 
Supernovae rates ${\cal R}$ in the rest frame are usually expressed in SNu,  
{\em i.e.} in supernovae per unit time and per unit blue luminosity  
($SNe / 10^{10} L_{\odot_B} / 100 yr$).  
The number of supernovae of a given type discovered during a search  
is related to the rate ${\cal R}$ of this type of SNe through  
\begin{equation} 
{\cal N} \ \sim\ \ {\cal R}\  
        \times\ \sum_{gal} L_{gal} \times T_{gal}  
\label{eq:snrate} 
\end{equation} 
where $L_{_{gal}}$ is the absolute luminosity of the galaxy $gal$, and  
$T_{gal}$ is the {\em control time} during which a SNIa could have  
been detected. If $\varepsilon(t,z,\ldots)$ is the search efficiency,  
{\em i.e.} the probability to detect a SNIa with redshift $z$ whose maximum  
occured at a time $t$ before the observation, $T_{gal}$ can be written  
as $T_{gal} = \int_{-\infty}^{+\infty} \varepsilon(t,z,\ldots) dt$.  
 
The sum ${\cal S} = \sum_{gal} L_{gal} \times T_{gal}$ is computed by 
Monte-Carlo integration. Firstly the galaxies in the search fields are 
detected using the program {\sf sextractor} (Bertin {\it et 
al.}~\cite{ber}).  Their apparent magnitudes are derived in the $R_c$ 
band from the EROS2 magnitudes. Since the redshift of each galaxy is 
not known, a value of $z$ is generated in the Monte-Carlo procedure, 
using a $p(z|R_{c_{gal}})$ pdf derived from the Schechter law with 
parameter values measured by the LCRS (Lin {\em et 
al.}~\cite{lin}). The absolute luminosities of each galaxy can then be 
calculated. The detection efficiency is fully simulated. The SN rate 
we thus obtain is 
\begin{equation} 
{\cal R} = 0.44^{+0.35\ +0.13}_{-0.21\ -0.07}\ h^2\ {\rm SNu}.  
\end{equation} 
 
By multiplying this value by the luminous density of the universe  
$\rho_L = 1.4 \pm 0.1\ 10^8\ h L_\odot Mpc^{-3}$ (Lin {\em et al.}~\cite{lin})  
we obtain the rate expressed in $h^3\ {\rm Mpc}^{-3}\ {\rm year}^{-1}$  
\begin{equation} 
{\cal R} = 0.62^{+0.49\ +0.19}_{-0.29\ -0.11} 10^{-4}\ h^3 {\rm Mpc}^{-3} {\rm year}^{-1}.  
\end{equation}

\begin{figure} 
\begin{center} 
\psfig{figure=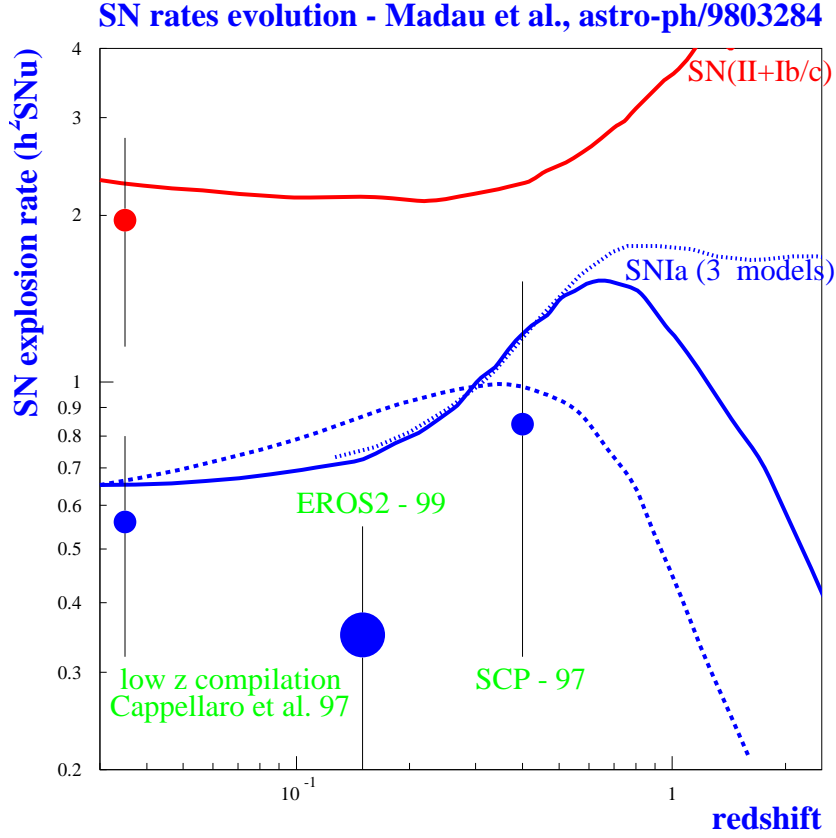,height=0.75\textwidth}  
\end{center} 
\vspace{-3mm} 
\caption{SNIa and SNII explosion rates as a function of redshift.  
(After Madau {\it et al.}$^9$) 
\label{fig:fgrate}} 
\end{figure}

\section*{Conclusion} 
  
 Since 1997, the EROS2 collaboration has conducted several campaigns 
 of supernovae searches. Our discovery rate is about 1 SN every two 
 hours of observations, which makes us competitive with respect to 
 other teams carrying on searches at the same $z$. In a first stage, 
 35 SNe were discovered, the light curves of 7 SNIa were studied, and 
 a first SNIa explosion rate at $z \sim 0.15$ was derived. In Spring 
 99, EROS2 participated in a worldwide search led by the {\em 
 Supernovae Cosmology Project}, and discovered 8 of the 19 SNIa near 
 maximum found by the consortium. Photometric and spectroscopic 
 follow-up data are currently been analyzed, and results are expected 
 to come out soon.

\end{document}